\documentclass[aps,prb,twocolumn,superscriptaddress,showpacs]{revtex4-1}
\usepackage{graphicx}
\usepackage{natbib}
\usepackage{color,ulem}
\usepackage{amsmath}
\usepackage{amssymb}
\usepackage{array}
\usepackage{multirow}

\begin{document}

\def\kc{KCu$_3$As$_2$O$_7$(OD)$_3$~}
\def\kcns{KCu$_3$As$_2$O$_7$(OD)$_3$}
\title{Helical order and multiferroicity in the $S=1/2$ quasi-kagome system \kc}


\author{G.~J.~Nilsen}
\email[Email address:~]{nilsen@ill.fr}
\affiliation{Institut Laue-Langevin, 6 rue Jules Horowitz, 38042 Grenoble, France}
\affiliation{Institute for Solid State Physics, University of Tokyo, Kashiwa, Chiba 277-8581, Japan}

\author{Y.~Okamoto}
\affiliation{Institute for Solid State Physics, University of Tokyo, Kashiwa, Chiba 277-8581, Japan}

\author{H.~Ishikawa}
\affiliation{Institute for Solid State Physics, University of Tokyo, Kashiwa, Chiba 277-8581, Japan}

\author{V.~Simonet}
\affiliation{Institut N\'{e}el, CNRS and Universit\'{e} Joseph Fourier, 38042 Grenoble, France}

\author{C.~V.~Colin}
\affiliation{Institut N\'{e}el, CNRS and Universit\'{e} Joseph Fourier, 38042 Grenoble, France}

\author{A. Cano}
\affiliation{CNRS, University of Bordeaux, ICMCB, UPR 9048, 33600 Pessac, France}

\author{L.~C.~Chapon}
\affiliation{Institut Laue-Langevin, 6 rue Jules Horowitz, 38042 Grenoble, France}

\author{T.~Hansen}
\affiliation{Institut Laue-Langevin, 6 rue Jules Horowitz, 38042 Grenoble, France}

\author{H.~Mutka}
\affiliation{Institut Laue-Langevin, 6 rue Jules Horowitz, 38042 Grenoble, France}

\author{Z.~Hiroi}
\affiliation{Institute for Solid State Physics, University of Tokyo, Kashiwa, Chiba 277-8581, Japan}

\date{\today}
\begin{abstract}
Several Cu$^{2+}$ hydroxide minerals have recently been identified as candidate realizations of the $S=1/2$ kagome Heisenberg model. In this context, we have studied the distorted system \kc using neutron scattering and bulk measurements. Although the distortion favors magnetic order over a spin liquid ground state, refinement of the magnetic diffraction pattern below $T_{N1}=7.05(5)$~K yields a complex helical structure with $\mathbf{k}=(0.77~0~0.11)$. This structure, as well as the spin excitation spectrum, are well described by a classical Heisenberg model with ferromagnetic nearest neighbor couplings. Multiferroicity is observed below $T_{N1}$, with an unusual crossover between improper and pseudo-proper behavior occurring at $T_{N2}=5.5$~K. The polarization at $T=2$~K is $P=1.5~\mu$Cm$^{-2}$. The properties of \kc highlight the variety of physics which arise from the interplay of spin and orbital degrees of freedom in Cu$^{2+}$ kagome systems.
\end{abstract}
\maketitle
The ground state of the highly frustrated $S=1/2$ kagome lattice Heisenberg model has inspired considerable debate over the last $20$ years, with recent theoretical candidates including exotic gapped $\mathcal{Z}_2$ liquid \cite{Yan2011} and gapless $U(1)$ spin liquid \cite{Iqbal2013} states. This has motivated an extensive search for potential realizations, which has thus far focused on Cu$^{2+}$-containing minerals, where the kagome lattices are made up of edge-sharing CuO$_6$ octahedra. A wide range of interesting magnetic behaviors have been observed in this family of materials; in herbertsmithite ($R\bar{3}m$) \cite{Shores2005}, no magnetic order and only short range dynamical correlations are found down to low $T$ \cite{Helton2007,Han2012,Vries2009}, while for volborthite ($C2/m$) \cite{Hiroi2001,Yoshida2009a,Nilsen2012,Yoshida2012a}, incommensurate long range order is only observed around $1$~K, despite $\theta_{CW} = -115$~K. In kapellasite \cite{Colman2008}, a polymorph of herbertsmithite, an unusual non-coplanar spin liquid is found to be promoted by disorder and further neighbor couplings \cite{Fak2012}. Numerous attempts have been made to understand this diversity theoretically; a symmetry analysis of the further neighbor Heisenberg model, for example, has identified $3$ possible coplanar and $5$ non-coplanar commensurate magnetically ordered ground states \cite{Messio2011}. Exact diagonalizations of finite clusters show that asymmetric exchange favors commensurate long range order \cite{Cepas2008,Rousochatzakis2009}, while depletion of the kagome lattice is understood to lead to a valence bond glass \cite{Singh2010}. 

The present study focuses on the recently uncovered synthetic Cu$^{2+}$-based mineral \kc \cite{Okamoto2012}, which, like volborthite, is monoclinically distorted. Elastic and inelastic neutron scattering experiments are combined with dielectric constant and polarization measurements to elucidate the unusual incommensurate magnetic ground state of \kcns, and identify it as the first multiferroic Cu$^{2+}$ kagome mineral.
\begin{figure}[th]
\centering
\includegraphics[width=0.75\linewidth]{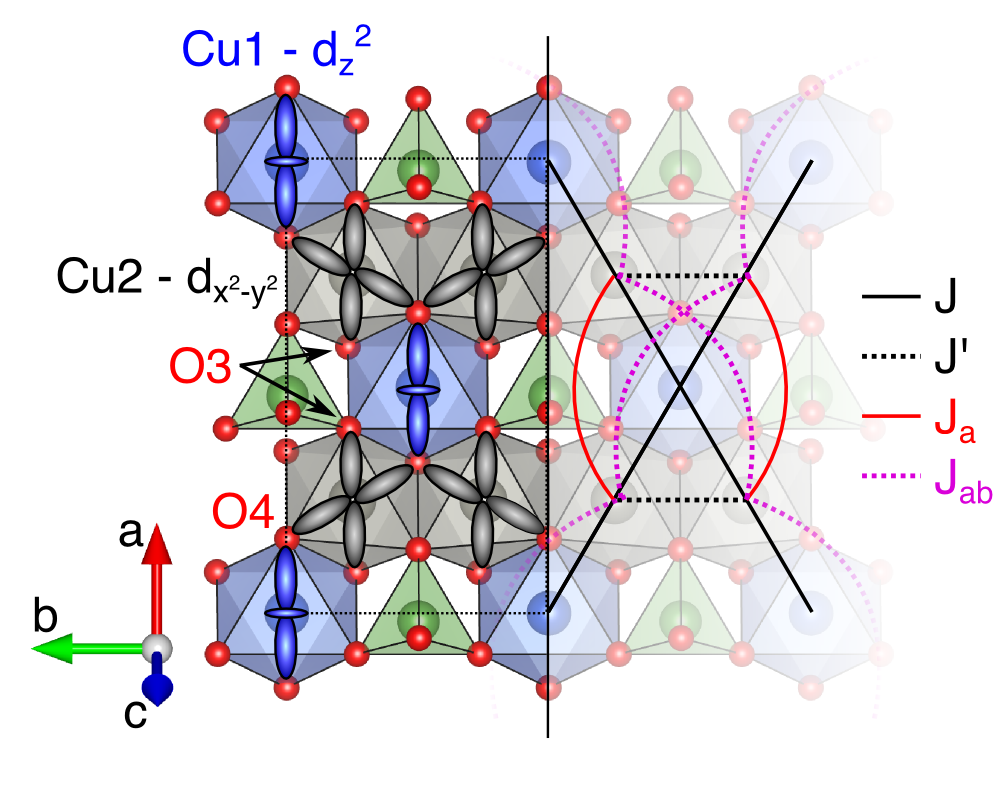}
\caption{(left) Structure and assumed orbital ordering pattern for \kcns. The kagome lattice, a 2D network of corner-sharing triangles, is formed by edge-sharing Cu1 (blue) and Cu2 (black) octahedra. (right) Magnetic exchange paths used in equation 1.}
\label{fig1}
\end{figure}

\kc crystallizes in the $C2/m$ space group, with lattice parameters $a=10.287$~\r{A}, $b=5.972$~\r{A}, $c=7.849$~\r{A}, and $\beta=117.74^\circ$  \cite{Effenberger1989}. There are two crystallographically inequivalent Cu$^{2+}$ sites: Cu1 at (0,0,0) (Wyckoff position $2a$) and Cu2 at (1/4,1/4,0) ($4e$). The CuO$_6$ octahedra share edges in the $ab$ plane, forming a distorted kagome lattice [Fig. \ref{fig1}]. An analysis of the distortions of the CuO$_6$ octahedra \cite{Okamoto2012} results in the assignment of $d_{x^2-y^2}$ as the magnetically active orbital on the Cu2 site, with $d_{z^2}$ singly occupied on Cu1. There are thus two types of nearest neighbor superexchange: neighboring Cu1 and Cu2 sites are connected by O4 in the $\mu_3$ bridging mode, with $\angle$Cu-O-Cu$=101.9^\circ$ and a Cu-Cu distance, $r_{\textnormal{Cu-Cu}}$, of $2.97$~\r{A}. The link between adjacent Cu2 is formed by the same O4, but with a slightly smaller $\angle$Cu-O-Cu$=101.4^\circ$, and longer $r_{\textnormal{Cu-Cu}}=2.99$~\r{A}. The exchange constants belonging to these pathways are referred to as $J$ and $J^\prime$, respectively (note the difference between this definition and that in [\onlinecite{Okamoto2012}]). 

In the conventional picture of superexchange across a $\mu_2$-bridging anion, the sign of $J$ is related to $\angle$Cu-O-Cu. For $\mu_3$-bridged Cu triangles, however, it is found that the sign is determined by the distance between $\mu_3$-O (here O4) and the plane of the Cu triangle (here $\Delta$Cu2-Cu2-Cu1), $r(\textnormal{O}-\textnormal{Cu}_3)$ \cite{Yoon2005}. In \kcns, $r(\textnormal{O}-\textnormal{Cu}_3)=0.85$~\r{A}, considerably in excess of the $r_c=0.35$~\r{A} beyond which a ferromagnetic $J$ is expected. Indeed, a Curie Weiss fit of the high temperature susceptibility yields a ferromagnetic Weiss constant $\theta_{CW} \sim 14$~K$\sim1.2$~meV \cite{Okamoto2012}. On the other hand, $r(\textnormal{O}-\textnormal{Cu}_3)>r_c$ also in vesignieite \cite{Okamoto2009} where $\theta_{CW}<0$. Unambigious assignment of the sign of superexchange based on structural parameters is therefore difficult in these materials. At lower temperatures, magnetic long range order is found to occur in \kc at $T_{N1}=7.1$~K, as indicated by a kink in $\chi(T)$ and a sharp anomaly in the specific heat $C_p(T)$. In addition, a shoulder is observed in the specific heat at $T_{N2}\simeq 5.5$~K, which we will later show to be another magnetic transition \cite{Okamoto2012}.    

\begin{figure}[th]
\centering
\includegraphics[width=0.53\linewidth]{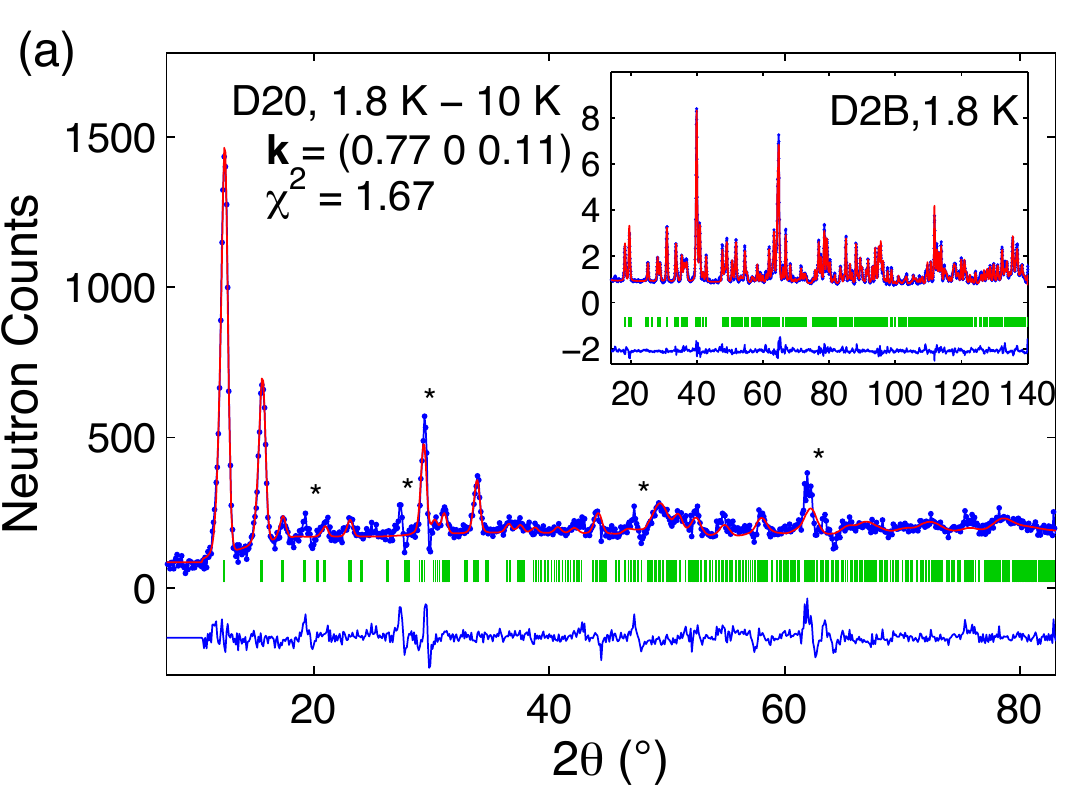}\includegraphics[width=0.47\linewidth]{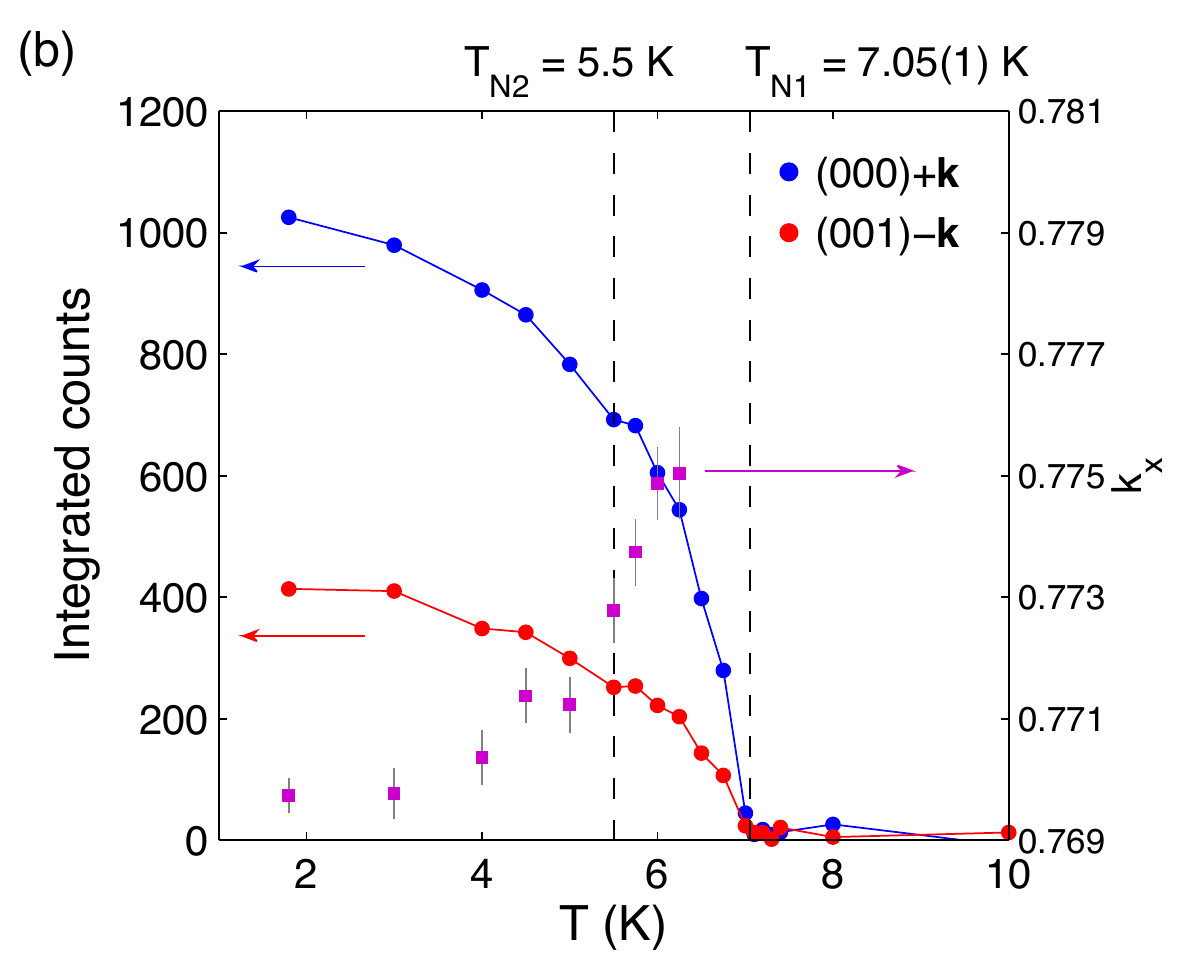}\\ \includegraphics[width=0.48\linewidth]{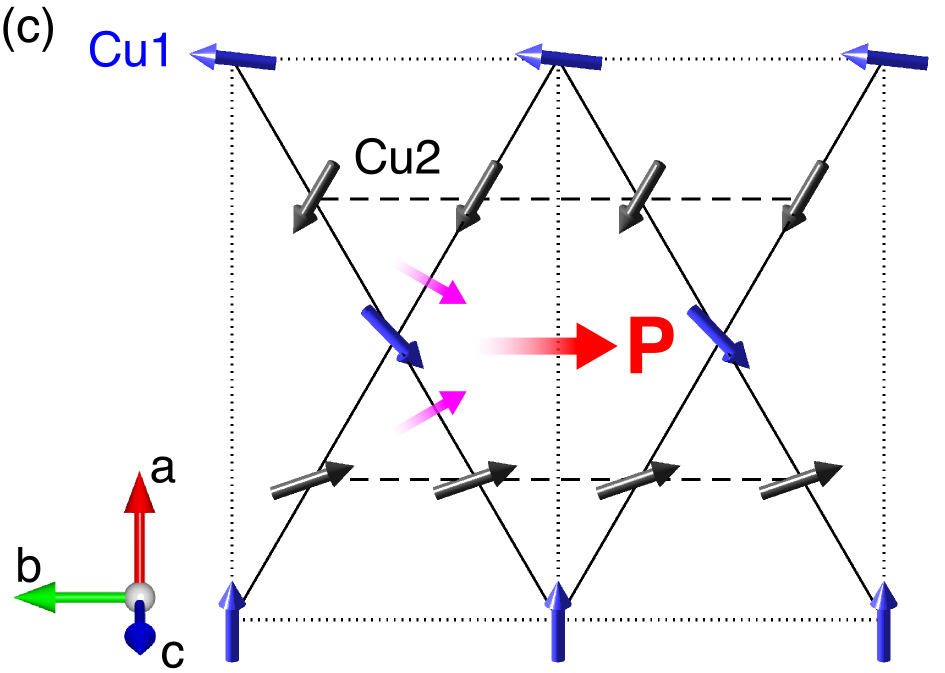} \includegraphics[width=0.48\linewidth]{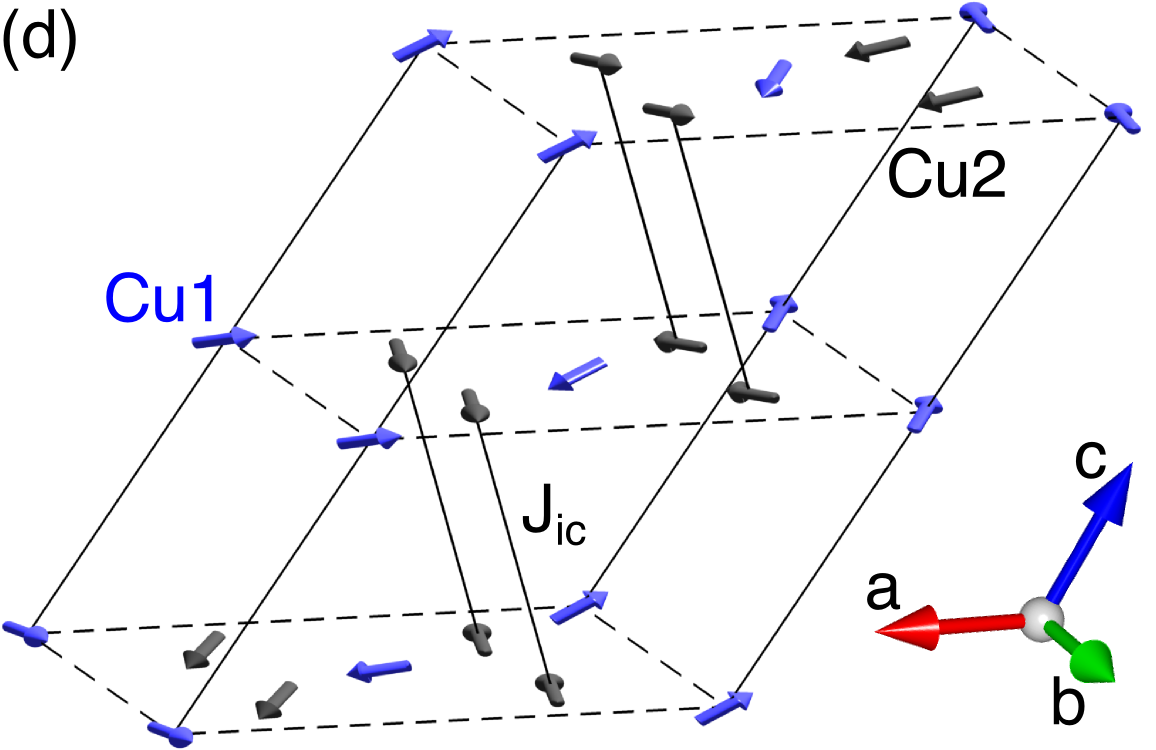}
\caption{(a) Subtracted elastic neutron scattering spectra (blue dots) from D20 with Rietveld refinement (red line) of the magnetic structure. The oscillations around nuclear Bragg positions (*) are an artefact resulting from the Debye-Waller factor. (inset) Rietveld refinement of the nuclear structure from high-resolution D2B data. (b) Magnetic intensity versus $T$ for the $(000)+\mathbf{k}$ and $(001)-\mathbf{k}$ reflections. The transitions observed in specific heat measurements are marked by dashed vertical lines. (c) The refined magnetic structure viewed along $\mathbf{c}^\ast$. The red and pink arrows indicate the total polarization and contributions to it from the $\left\langle110\right\rangle$ chains, respectively. (d) The propagation of the magnetic structure along $c$. Note the antiparallel alignment of spins related by the centering translation on adjacent planes. The Cu-Cu pathway between these is also the shortest, and is thus assigned to $J_{ip}$.}
\label{fig2}
\end{figure}

A deuterated sample suitable for neutron scattering experiments was synthesized by performing the previously reported synthesis in D$_2$O rather than H$_2$O solvent \cite{Okamoto2012}. All operations were carried out in a dry atmosphere to avoid exchange between D and H, and the sample purity was verified by both X-ray diffraction and SQUID magnetometry. Around $5$~g of powder sample were loaded in a $9$~mm V can, and cooled to temperatures down to $1.8$~K using a standard $^4$He Orange cryostat. Neutron diffraction patterns were collected on the D20 instrument at ILL using $\lambda=2.41$~\r{A} neutrons from a PG$(002)$ monochromator at takeoff angle $\theta=42^\circ$; these conditions optimize flux at the expense of resolution. The background from the sample environment was reduced using a radial oscillating collimator, and spectra were collected at $T=1.8<T_N<10$~K. Higher resolution patterns for determination of the nuclear structure were measured at $T=1.8$~K on D2B (ILL), using $\lambda=1.594$~\r{A} from the $(335)$ reflection of a Ge monochromator with $\theta=135^\circ$. Inelastic neutron scattering data were collected at the IN4 (ILL) time-of-flight spectrometer with an incident energy $E_i=9.2$~meV ($\lambda=2.981$~\r{A}). Measurements of the dielectric constant, $\epsilon_r$, were carried out on a pressed pellet of thickness $d=0.65$~mm silver pasted to flat electrodes, and connected to an LCR meter (Agilent E4980A). Finally, the polarization, $P$, was determined by integrating the pyroelectric current measured on the same pellet using an electrometer (Keithley 6517A), cooling in electric fields of $\pm307$~kVm$^{-1}$.

We begin our discussion of our results by establishing the nuclear structure; a Rietveld refinement of the high resolution diffraction data yields lattice parameters $a=10.2872(1)$~\r{A}, $b=5.9728(1)$~\r{A}, $c=7.8492(1)$~\r{A}, and $\beta=117.740(1)^\circ$, which agree well with those reported in [\onlinecite{Effenberger1989}]. The H/D atoms are placed based on symmetry considerations, and refinement yields coordinates $(0.261(1),0,0.265(1))$ for H1 and $(0.5,0,0.5)$ for H2, respectively. H2 forms a hydrogen bond along with two O2 atoms belonging to adjacent AsO$_4$ groups. The deuteration was estimated to be $96.0(2)$~\% on the H1 site and $84(2)$~\% on H2. 

In the medium resolution D20 data, several magnetic Bragg peaks are observed below $T_{N1}=7.05(5)$~K, all of which may be uniquely indexed by the incommensurate $\mathbf{k}$-vector $=(0.7750(4)~0~0.1090(4))$ at $T=6.25$~K. Cooling further, $k_x$ ($k_z$) decreases (increases), saturating at $\mathbf{k}=(0.7697(1)~0~0.1109(1))$ at $1.8$~K [Fig. 2(b)]. Around $T_{N2}=5.5$~K, there is a discontinuity in the intensities of the $(000)+\mathbf{k}$ and $(001)-\mathbf{k}$ Bragg peaks, and an accompanying small shift in their relative intensity. This overall temperature dependence is compatible with the thermodynamic data \cite{Okamoto2012}. As no additional Bragg peaks are observed below $T_{N2}$, the subtle difference in magnetic structure between $T_{N2}<T<T_{N1}$ and $T<T_{N2}$ is not distinguishable in the present study. Accordingly, we focus on the $T=1.8$~K$<T_{N2}$ data for the subsequent analysis of the magnetic structure.

Because the experimentally determined $\mathbf{k}$-vector does not coincide with a high symmetry point in the Brillouin zone, a full description of the magnetic symmetry requires determining the irreducible corepresentations of the magnetic group $\mathbf{M}=\mathbf{G}^k+K\mathbf{h}$, where $\mathbf{S}=\{E,m_y,2_y,\bar{1}\}$ is the space group, $\mathbf{G^k}=\{E,m\}$ is the little group of operators leaving $\mathbf{k}$ invariant, $\mathbf{h}=\mathbf{S}-\mathbf{G^k}$, and $K$ is the complex conjugation operator  \cite{Radaelli2007}. We thus find two one-dimensional real coreps, $D_1$ and $D_2$, for which the basis functions $\psi_{\alpha}$ may be calculated by the usual projection method. The $D_1$ mode has the spins on the Cu1 site pointing in the $b$ direction, with the $a$ and $c$ components antiparallel between the two orbits of the Cu2 site (Table 1). For the $D_2$ mode, on the other hand, the magnetic moment on the Cu1 site lies in the $ac$ plane, while the $b$ components of the Cu2 spins are antiparallel.

\begin{table}
\caption{\label{table1} Characters and basis functions $\psi_\alpha$ of the coreps of the magnetic group $\mathbf{M}$. The Cu2 site is split into two orbits, with Cu2$^\prime$ at $(1/4,3/4,0)$.}
\begin{ruledtabular}
\footnotesize
\begin{tabular}{*{6}{l}}
\multirow{2}{*}{Corep/Irrep} & $E$ & $2_y$ & \multirow{2}{*}{$\psi_{Cu1}$} & \multirow{2}{*}{$\psi_{Cu2}$} & \multirow{2}{*}{$\psi_{Cu2^\prime}$} \\
 & $K\bar{1}$ & $Km_y$ &  &  &    \\
\hline
$D_1/\Gamma_1$ & 1 & 1 & (0,1,0) & (1,1,1) & (-1,1,-1)  \\
$D_2/\Gamma_2$ & 1 & -1 & (1,0,1) & (1,1,1) & (1,-1,1) \\
\end{tabular}
\end{ruledtabular}
\end{table}

Fitting the magnetic scattering, $I_{mag}=I(1.8~\textnormal{K})-I(10~\textnormal{K})$, to either of the above modes individually yields poor agreement with the experimental data; $\chi^2>4$ in both cases. We must therefore shift our attention to solutions involving both. Considering only solutions where the modes are summed in quadrature and recalling that the elements of $\mathbf{h}$ complex conjugate the basis functions, two possible types of state result; a helix with point group $21^\prime$ for $D_1+iD_2$ ($iD_1+D_2$) and an amplitude modulated structure with point group $m1^\prime$ for $D_1+iD_1$ ($D_2+iD_2$). Because the structure is defined by a single $\mathbf{k}$-vector, we employ the refinement constraint that the spins on either sub-lattice must be coplanar. In addition, for the helical $D_1+iD_2$-type structure, we restrict the envelope of the helix to being circular by refining the spherical components $(\mu,\theta,\phi)$ of the Fourier coefficients in the $A2/m$ setting of the space group ($a^\prime=c,~b^\prime=a,~c^\prime=b$). We note that the point groups of both possible solutions are polar, and therefore compatible with ferroelectricity. 

The best fit ($\chi^2=1.67$) is achieved for the $iD_1+D_2$ solution, with the plane of the helix rotated out of the $ac$ plane by the azimuthal angle $\phi_{Cu1}=\phi_{Cu2}=148(2)^\circ$ and the polar angle $\theta_{Cu1}=\theta_{Cu2}=166(5)^\circ$. The Cu2 ($4e$) site is split into two orbits with the same phase  $\phi_{Cu2}=\phi_{Cu2^\prime}=k_x/4=0.1923$ [Fig. 2(a)]. The ordered moments are found to be $0.86(2)~\mu_B$ for Cu1 and $0.87(1)~\mu_B$ for Cu2, close to the full $1\mu_B$ expected for Cu$^{2+}$, suggesting a surprisingly small fluctuating component for a frustrated quantum spin system. The magnetic structure may be described as crossing helical chains along the $\langle110\rangle$ directions, with ferromagnetic alignment along the $b$-direction [Fig. \ref{fig2}(c)]. Along $c$, the moment direction is slowly modulated, with a period of approximately $10$ unit cells [Fig. \ref{fig2}(d)]. Spins related to each other by the centering translation are antiparallel on adjacent layers; \textit{i.e.} $\mathbf{m}(1/4,1/4,0)=-\mathbf{m}(3/4,1/4,1)$. Finally, the plane of rotation for both sites is tilted out of the $ab$-plane by $31(2)^\circ$.

\begin{figure}[th]
\centering
\includegraphics[width=1\linewidth]{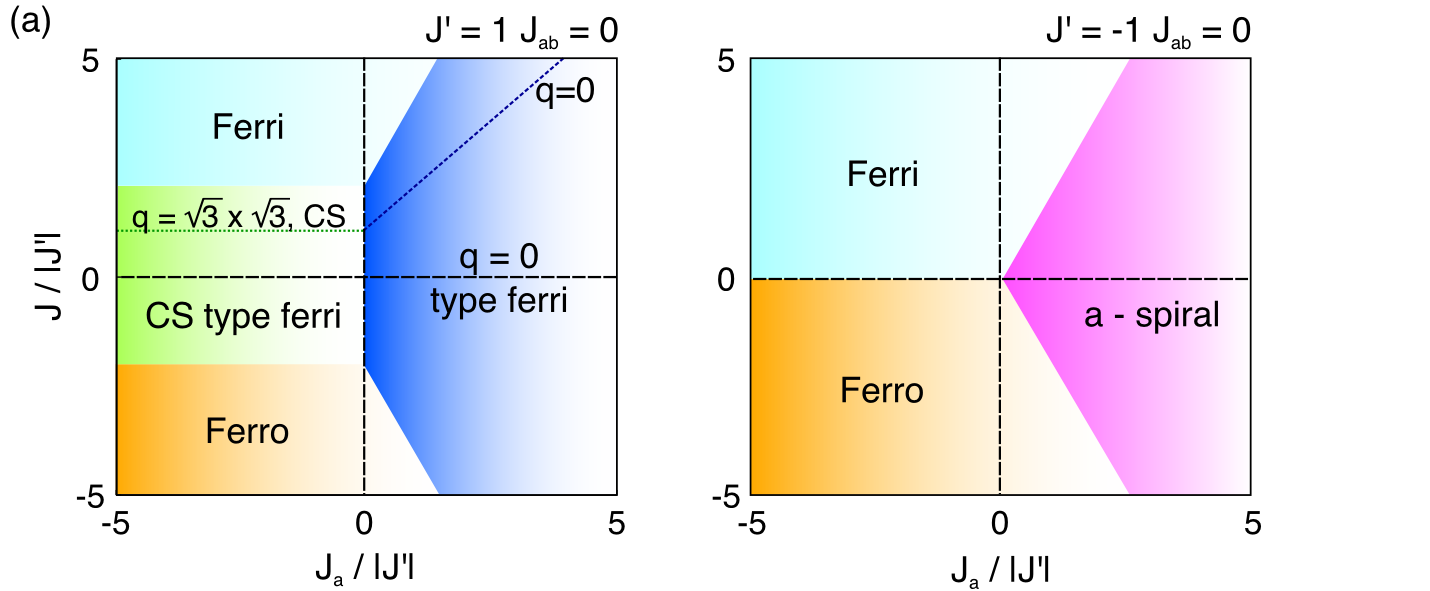} \\
\includegraphics[width=1.0\linewidth]{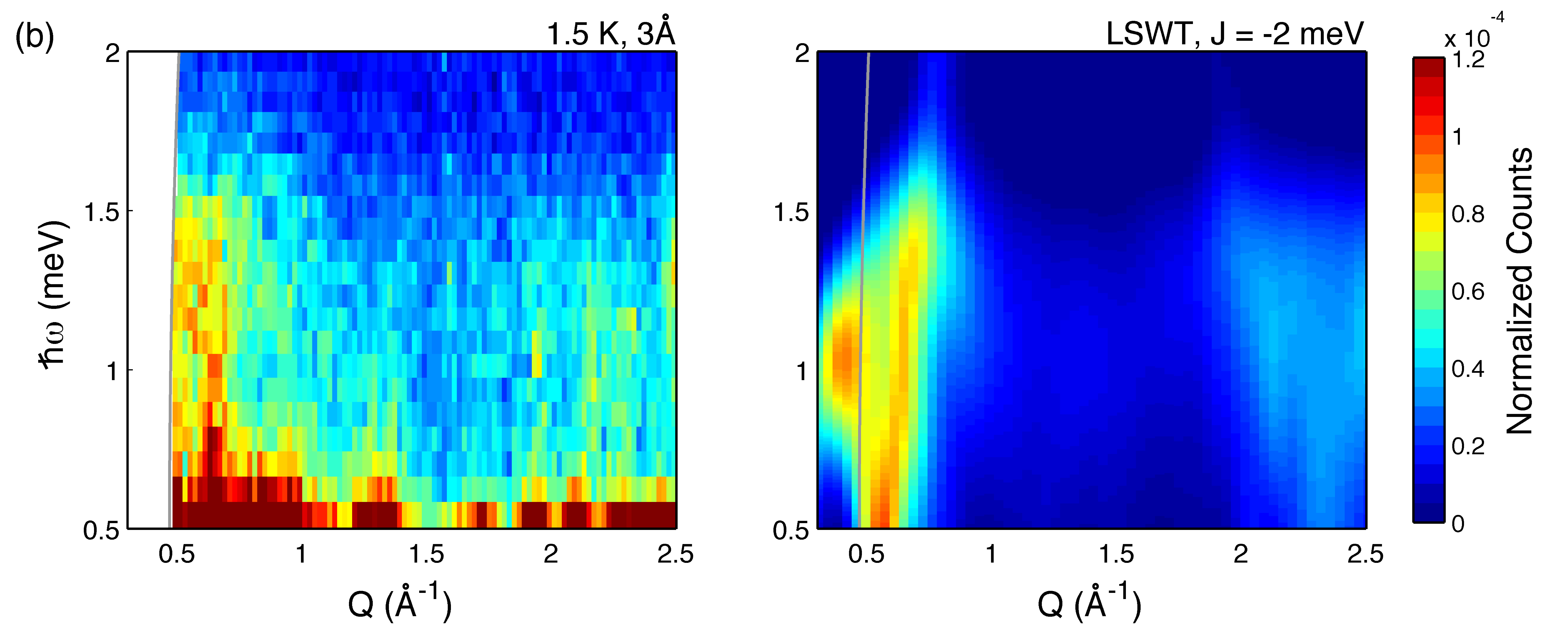}
\caption{(a) Phase diagrams in the $J^\prime=1$ and $J^\prime=-1$ planes derived by solving equation 1 subject to equation 2. When $J^\prime<0$, a large area is occupied by the experimentally observed $a$-helical phase. (b) Comparison of linear spin wave theory (right) and inelastic neutron scattering data (left) for the parameters listed in the text.}
\label{fig4}
\end{figure}

Next, we seek a microscopic model that explains the refined magnetic structure. It is clear that the nearest neighbor Heisenberg Hamiltonian on the kagome lattice is insufficient for this purpose. As such, the presence of either further neighbor couplings or anisotropic terms in the Hamiltonian are required. We identify two likely further neighbor superexchange pathways within the kagome planes: one between Cu2 atoms along the $\langle 100 \rangle$ direction, mediated by two O3 atoms [Fig. 1], with $2(\angle$Cu-O-O$)=187.2^\circ$ and $r$(Cu-Cu) = 5.12~\r{A}, and the other along $\langle 110 \rangle$ through O3 and O4, with $2(\angle$Cu-O-O$) = 291.8^\circ$. The latter is similar to the pathway which yields antiferromagnetic next nearest neighbor exchange in edge-shared square planar Cu$^{2+}$ systems like LiCuVO$_4$ \cite{Enderle2005} and CuGeO$_3$ \cite{Hase1993}. The exchange integrals corresponding to the two pathways above will be referred to as $J_a$ and $J_{ab}$, respectively. If $J_a>0$, as seems likely from $\angle$Cu-O-O-Cu$\sim180^\circ$, it frustrates the nearest neighbor coupling $J$ when $J<0$ [Fig. \ref{fig1}]. Likewise, for $J_{ab}>0$, frustration arises when $J^\prime<0$. An interplane coupling is also required to explain the modulation along the $c$-direction; the only plausible pathway is through the AsO$_4$ tetrahedra, which are joined by H(2). The simplest model which can be constructed for Heisenberg spins is then:
\begin{align}
\mathcal{H}=\sum_{i,j} J(\left| \mathbf{r}_i - \mathbf{r}_j \right|) \mathbf{S}_i\cdot\mathbf{S}_j,
\label{eqn1}
\end{align}
where the sum is over all pairs of classical unit spins $\mathbf{S}_i,\mathbf{S}_j$ connected by $J(\left| \mathbf{r}_i - \mathbf{r}_j \right|)$.

Since the experimentally determined structure is a coplanar helix, the spin on each sublattice may be written;
\begin{align}
\mathbf{S}_i^{\alpha}=S\left[\hat{x}\cos{\left(\mathbf{k}\cdot\mathbf{r}_i+\phi_{\alpha}\right)}-\hat{y}\sin{\left( \mathbf{k}\cdot\mathbf{r}_i+\phi_{\alpha}\right)} \right],
\end{align}
where $\hat{x}$ and $\hat{y}$ are unit vectors along $a$ and $b$, and $\phi_{\alpha}$ is the phase on sublattice $\alpha$. Setting $J_{ab}=0$, $J^{\prime}=\pm1$, $J_{ip}=0.01$, and minimizing with respect to $\mathbf{k}$ and $\phi^\alpha$, we thus arrive at the phase diagrams shown in figure $3(a)$. In the $J^\prime=1$ plane, there are, in addition to the trivial ferromagnetic and N\'{e}el ferrimagnetic states, two tilted ferrimagnetic states interpolating between them, as well as narrow strips of uniform ($\mathbf{q}=0$), striped (CS), or staggered chirality order ($\mathbf{q}=\sqrt{3}\times\sqrt{3}$). For $J^\prime=-1$, the phase diagram is considerably simpler, and for antiferromagnetic $J_a$, the $a$-helix is favored. The components of the $\mathbf{k}$-vector in this state are found to be $k_x=2-2/\pi[\cos^{-1}\left(|J|/J_a)\right)]$ and $k_z=(1-k_x)/2$ for antiferromagnetic $J_{ip}$. The experimental $\mathbf{k}$ is thus accurately reproduced for the condition $|J|\sim 0.68J_a$, for $J,J^\prime<0$; when $J_{ab}$ is included, the right side of the previous expression becomes $0.68(J_{a}+J_{ab})$.

To gain a more quantitative grasp of the magnitudes of $J(\left| \mathbf{r}_i - \mathbf{r}_j \right|)$, we proceed to calculate the linear spin wave spectrum for comparison with $S(Q,\hbar \omega)$ collected in the ordered phase on the inelastic neutron spectrometer IN4 [Fig. 3]. In addition to the requirement determined above, another constraint on $J(\left| \mathbf{r}_i - \mathbf{r}_j \right|)$ arises from the Weiss constant, $\theta_{CW}=-(4k_B)^{-1}\sum_{i=1}^{z}{z_iJ_i}\sim1.3$~meV. Solutions with both ferromagnetic and antiferromagnetic $J$ were calculated, but only those with both $J$ and $J^\prime$ ferromagnetic are able to generate the correct bandwidth whilst simultaneously giving a reasonable $\theta_{CW}$. In addition, $J_a>J_{ab}$ is required to reproduce the reduction in intensity above $\hbar\omega=1.5$~meV. The powder averaged $S(Q,\hbar\omega)$ for the following parameter set is shown in figure 3(b):  $J=-2$~meV, $J^\prime=-1$~meV, $J_{ab}=0.75$~meV, $J_a=2.2$~meV, $J_{ip}=0.01$~meV. Beyond the good qualitative agreement with experiment, this solution gives an ordered moment $\sim0.8\mu_B$ on both sites, consistent with the refined moment. 

\begin{figure}[th]
\centering
\includegraphics[width=\linewidth]{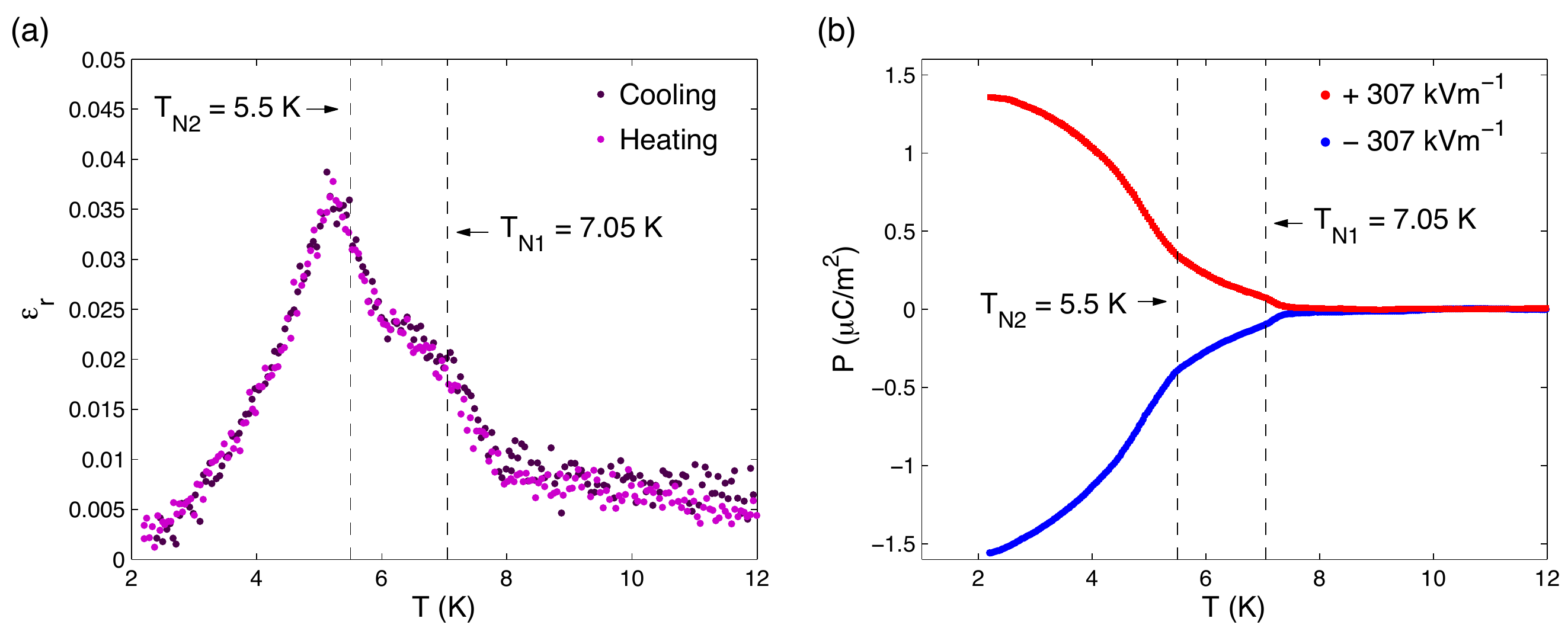}
\caption{(a) The temperature dependence of the dielectric constant, $\epsilon_r$, measured at $10$~kHz with Vac=1~V. A linear background has been subtracted for clarity. Anomalies are observed at both $T_{N1}$ and $T_S$. (b) The polarization, $P$, versus temperature. The red (blue) curve corresponds to a sample poled in an electric field of $+307~(-307)$~kVm$^{-1}$. In both panels, the features observed by other probes at $T_{N1}$ and $T_S$ are marked by dashed vertical lines.}
\label{fig3}
\end{figure}

While the further-neighbor Heisenberg model proposed above has thus far been successful in describing our experimental data, it should be noted that Dzyaloshinskii-Moriya (DM) couplings are also allowed in \kcns. Their presence is an important difference with respect to other helical magnets, such as LiCuVO$_4$, where this type of interaction is forbidden by symmetry \cite{Mourigal2011}. Given that DM terms favor orthogonal alignment of spins, and therefore alter the pitch angle of the helix, their experimental determination is particularly important.

We finally move on to the magneto-electric properties of \kcns; according to the $21^\prime$ point-group symmetry of the low-temperature helical magnetic structure, an electric polarization is anticipated along the $b$-direction. Indeed, we observe a switchable spontaneous polarization approaching $P\sim 1.5~\mu$Cm$^{-2}$ at $2$~K, which is compatible with ferroelectricity of magnetic origin \cite{Cheong2007}. The temperature dependence of this polarization reveals a number of remarkable features: firstly, $P$ becomes non-zero at $T_{N1}$, below which it increases linearly [Fig. 4(b)]. This, along with the step-like anomaly in $\epsilon_p$ at $T_{N1}$ [Fig 3.(a)], is a hallmark of improper ferroelectricity, as originally defined in [\onlinecite{Levanyuk1974}]. The linear $T$-dependence reveals a coupling $P\eta^2$ with the primary order parameter, $\eta$, associated with the magnetic order \footnote{We note that neither corep $D_1$ nor $D_2$ are polar. The ferrolectricity observed below $T_{N1}$ thus implies that both participate in the magnetic ordering and, consequently, that the transition should be discontinuous. This is in contradiction with the second-order transitions observed experimentally [Figs. 2(b) and 4], which suggests that spin-orbit coupling is marginal and the transitions are better described within the exhange approximation. In this case, the order parameter belongs to the exchange multiplet $\eta = D_1+2D_2$ of representations that remain invariant under spin rotation \cite{Izyumov1979}.}. At $T_{N2}$, a transformation to more conventional (pseudo-proper) ferroelectric behavior is observed, with a divergent dielectric constant and a square-root behavior for the electric polarization. This change is absent in model spin-driven ferroelectrics like TbMnO$_3$ \cite{Kimura2003} and TbMn$_2$O$_5$ \cite{Hur2004}, which display only the pseudo-proper regime \cite{Toledano2009,Cano2010}. The striking concurrence of ferroelectricity with magnetic ordering and its transformation from improper to pseudo-proper suggests that the multiferroic interplay in kagome systems is both complex and rich. Its understanding requires a more precise characterization of the order that appears below $T_{N1}$, which will be reported elsewhere.

In conclusion, we have carried out elastic and inelastic neutron scattering experiments, as well as dielectric and polarization measurements on the quasi-kagome $S=1/2$ quantum magnet \kcns, 
These experiments convincingly reveal (i) helical magnetic order with incommensurate propagation vector $\mathbf{k}=(0.77,~0,~0.11)$ and (ii) well-defined spin-wave excitations, both explained by a Heisenberg exchange model. In addition, the system displays (iii) spin-driven ferroelectricity that appears at the onset of magnetic order and transforms from improper to pseudo-proper at lower temperature. \kc is, to our knowledge, the first $S=1/2$ kagome lattice system to show multiferroicity.

\begin{acknowledgments}
We thank S\'{a}ndor T\'{o}th for useful discussions on the spin wave simulations and for sharing the SpinW code \cite{Toth2014}, Ludovic Gendrin and St\'{e}phane Rols for support during the neutron experiments, and F. Gay for his help in developing the dielectric measurements set up. GJN is grateful for the grant of a visiting professorship from the ISSP, during which this article was written.
\end{acknowledgments}

\bibliographystyle{apsrev4-1}
\bibliography{kas}
 
\end{document}